\documentclass[twocolumn,showpacs,amsmath,amssymb,prl, floatfix]{revtex4}
\preprint{Draft-PRL}
\usepackage{graphicx}
\usepackage{dcolumn}
\usepackage{bm}

\newcommand{\etap}{\eta^\prime}
\newcommand{\phet}{$J/\psi\to\phi\eta$}
\newcommand{\phett}{$J/\psi\to\phi\eta(\eta^\prime)$}

\begin{document}

\title{Search for Invisible Decays of $\eta$ and $\etap$ in
  $J/\psi  \rightarrow \phi\eta$ and $\phi \etap$}

\author{
  M.~Ablikim$^{1}$, J.~Z.~Bai$^{1}$, Y.~Ban$^{12}$, J.~G.~Bian$^{1}$,
  X.~Cai$^{1}$, H.~F.~Chen$^{17}$, H.~S.~Chen$^{1}$, H.~X.~Chen$^{1}$,
  J.~C.~Chen$^{1}$, Jin~Chen$^{1}$
% ...
  \\
  \vspace{0.2cm}
  (BES Collaboration)\\
  \vspace{0.2cm} {\it
    $^{1}$ Institute of High Energy Physics, Beijing 100049, People's Republic of China\\
% ...
$^{12}$ Peking University, Beijing 100871, People's Republic of China\\
% ...
$^{17}$ University of Science and Technology of China, Hefei 230026, People's Republic of China\\
% ...
} }

\date{\today}

\begin{abstract}
  Using a data sample of $58\times 10^6$ $J/\psi$ decays collected
  with the BES II detector at the BEPC, searches for invisible decays
  of $\eta$ and $\eta^\prime$ in $J/\psi$ to $\phi\eta$ and
  $\phi\etap$ are performed. The $\phi$ signals, which are
  reconstructed in $K^+K^-$ final states, are used to tag the $\eta$ and
  $\eta^\prime$ decays. No signals are found for the invisible decays
  of either $\eta$ or $\eta^\prime$, and upper limits at the 90\%
  confidence level are determined to be $1.65 \times 10^{-3}$ for the
  ratio $\frac{B(\eta\to \text{invisible})}{B(\eta\to\gamma\gamma)}$
  and $6.69\times 10^{-2}$ for $\frac{B(\etap\to
  \text{invisible})}{B(\etap\to\gamma\gamma)}$. These are the first
  searches for $\eta$ and $\eta^\prime$ decays into invisible final
  states.
\end{abstract}


\maketitle

Invisible decays of quarkonium states such as the $J/\psi$ and the
$\Upsilon$, etc., offer a window into what may lie beyond the Standard
Model (SM)~\cite{Fayet:1979qi,Fayet:2006sp}. The reason is that apart
from neutrinos, the Standard Model includes no other invisible final
particles that these states can decay into.  It is such a window that
we intend to further explore by presenting here the first experimental
limits on invisible decays of the $\eta$ and $\eta'$, which complement
the limit of $2.7 \ 10^{-7}$ recently established in
\cite{Artamonov:2005cu} for the invisible decays of the $\,\pi^\circ$.

Theories beyond the SM generally include new physics, such as,
possibly, light dark matter (LDM) particles~\cite{Boehm:2003hm}. These
can have the right relic abundance to constitute the nonbaryonic dark
matter of the Universe, if they are coupled to the SM through a new
light gauge boson $U$~\cite{Fayet:1980ad}, or exchanges of heavy
fermions.  It is also possible to consider a light neutralino with
coupling to the SM mediated by a light scalar singlet in the
next-to-minimal supersymmetric standard model~\cite{Ellis:1988er}.

Recently, observations of a bright 511 keV $\gamma$-ray line from the
galactic bulge have been reported by the SPI spectrometer on the
INTEGRAL satellite~\cite{Jean:2003ci}.  The corresponding galactic
positron flux, as well as the smooth symmetric morphology of the 511
keV emission, may be interpreted as originating from the annihilation
of LDM particles into $e^+e^-$ pairs ~\cite{Boehm:2003hm} (also
constrained by \cite{Beacom:2004pe}).  It is in any case very
interesting to search for such light invisible particles in collider
experiments.  CLEO gave an upper bound on $\Upsilon(1S) \rightarrow
\gamma + \text{invisible}$, which is sensitive to dark matter
candidates lighter than about 3 GeV/$c^2$~\cite{Balest:1994ch}, and
also provides an upper limit on the axial coupling of the new $U$
boson to the $b$ quark.  It is crucial, in addition, to search for the
invisible decays of light quarkonium ($q\overline{q}$, $q= u$,$d$, or
$s$ quark) states which can be used to constrain the masses of LDM
particles and the couplings of the new boson to the light
quarks~\cite{Fayet:2006sp}. We present here measurements of branching
fractions of $\eta$ and $\eta^{\prime}$ decays into invisible final
states.

The data used in this analysis, consisting of $58 \times 10^6$
$J/\psi$ events, were accumulated with the BES II
detector~\cite{Bai:1994zm}, at the BEPC.  BES II is a conventional
solenoidal magnetic detector that is described in detail in
Ref.~\cite{Bai:1994zm}.  A 12-layer vertex chamber (VC) surrounding
the beam pipe provides trigger and coordinate information. A
forty-layer main drift chamber (MDC), located radially outside the VC,
provides trajectory and energy loss (dE/dx) information for charged
tracks over 85\% of the total solid angle. The momentum resolution is
$\sigma _p/p = 0.017 \sqrt{1+p^2}$ ($p$ in $\hbox{\rm GeV}/c$), and
the $dE/dx$ resolution for hadron tracks is $\sim 8\%$.  An array of
48 scintillation counters surrounding the MDC measures the
time-of-flight (TOF) of charged tracks with a resolution of $\sim 200$
ps for hadrons.  Radially outside the TOF system is a 12 radiation
length, lead-gas barrel shower counter (BSC).  This measures the
energies of electrons and photons over $\sim 80\%$ of the total solid
angle with an energy resolution of $\sigma_E/E=22\%/\sqrt{E}$ ($E$ in
GeV).  Outside of the solenoid coil, which provides a 0.4~Tesla
magnetic field over the tracking volume, is an iron flux return that
is instrumented with three double layers of counters that identify
muons with momentum greater than 0.5~GeV/$c$.

\begin{figure}[htbp]
  \centering
  \includegraphics[width=0.35\textwidth,height=0.15\textheight]{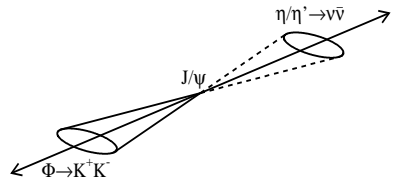}
  \caption{Schematic of $J/\psi \rightarrow \phi \eta$ or $\phi
    \eta^\prime$. The $\phi$, which is reconstructed in $K^+K^-$ final
    states, can be used to tag the invisible decay of the $\eta$ and
    $\eta^{\prime}$.}
  \label{fig:demo}
\end{figure}

In order to detect invisible $\eta$ and $\eta^{\prime}$ decays, we use
$J/\psi \rightarrow \phi\eta$ and $\phi\eta^{\prime}$ decays.  These
two-body decays provide a very simple event topology, as shown in
Fig.~\ref{fig:demo}, in which the $\phi$ signals can be reconstructed
easily and cleanly decaying into $K^+ K^-$.  The reconstructed $\phi$
particles can be used to tag the $\eta$ and $\eta^{\prime}$ in order
to allow a search for their invisible decays.  Since both $\phi$ and
$\eta$ ($\eta^{\prime}$) have narrow widths, which are negligible
compared with the detector resolution, the shape of the momentum
distribution of the $\phi$ is approximately Gaussian.  The mean value
of the $\phi$ momentum distribution is 1.320 GeV/$c$ for \phet \mbox{}
and 1.192 GeV/$c$ for $J/\psi\to\phi\eta^\prime$.  The missing
momentum, ${P}_{miss} = |\vec{P}_{miss}|$, is a powerful
discriminating variable to separate signal events from possible
backgrounds, in which the missing side is not from $\eta$
($\eta^{\prime}$) decay. Here, $\vec{P}_{miss} = - \vec{P}_{\phi}$.
The $\eta$ and $\eta^{\prime}$ signal regions are defined as
$|P_{miss} - 1.320| < 3 \sigma^{\eta}_{reso}$ for \phet \mbox{} and
$|P_{miss} - 1.192| < 3 \sigma^{\eta^{\prime}}_{reso}$ for
$J/\psi\to\phi\eta^\prime$, where $\sigma^{\eta}_{reso}$ (22 MeV/$c$)
and $\sigma^{\eta^{\prime}}_{reso}$ (20 MeV/$c$) are detector
resolutions of $P_{miss}$ for \phet \mbox{} and
$J/\psi\to\phi\eta^\prime$, respectively.  In addition, the $\eta$ and
$\eta^{\prime}$ decay regions are easy to define in the lab system due
to the strong boost of the $\phi$ from $J/\psi$ decay, as shown in
Fig.~\ref{fig:demo}.

In the event selection, the total number of charged tracks is required
to be two with net charge zero. Each track should have a good helix
fit in the MDC, and the polar angle $\theta$ must satisfy
$|\cos\theta| < 0.8$. The event must originate near the collision
point; tracks must satisfy $\sqrt{x^2 + y^2} \le 2$ cm, $|z| \le 20$
cm, where $x$, $y$, and $z$ are the space coordinates of the point of
closest approach of tracks to the beam axis. Particle identification
(PID) is performed using combined TOF and $dE/dx$ information, and
both charged tracks must be identified as kaons.

\begin{figure}[htbp]
  \centering
  \begin{minipage}{1.0\linewidth}
  \includegraphics[width=\textwidth,height=0.18\textheight]{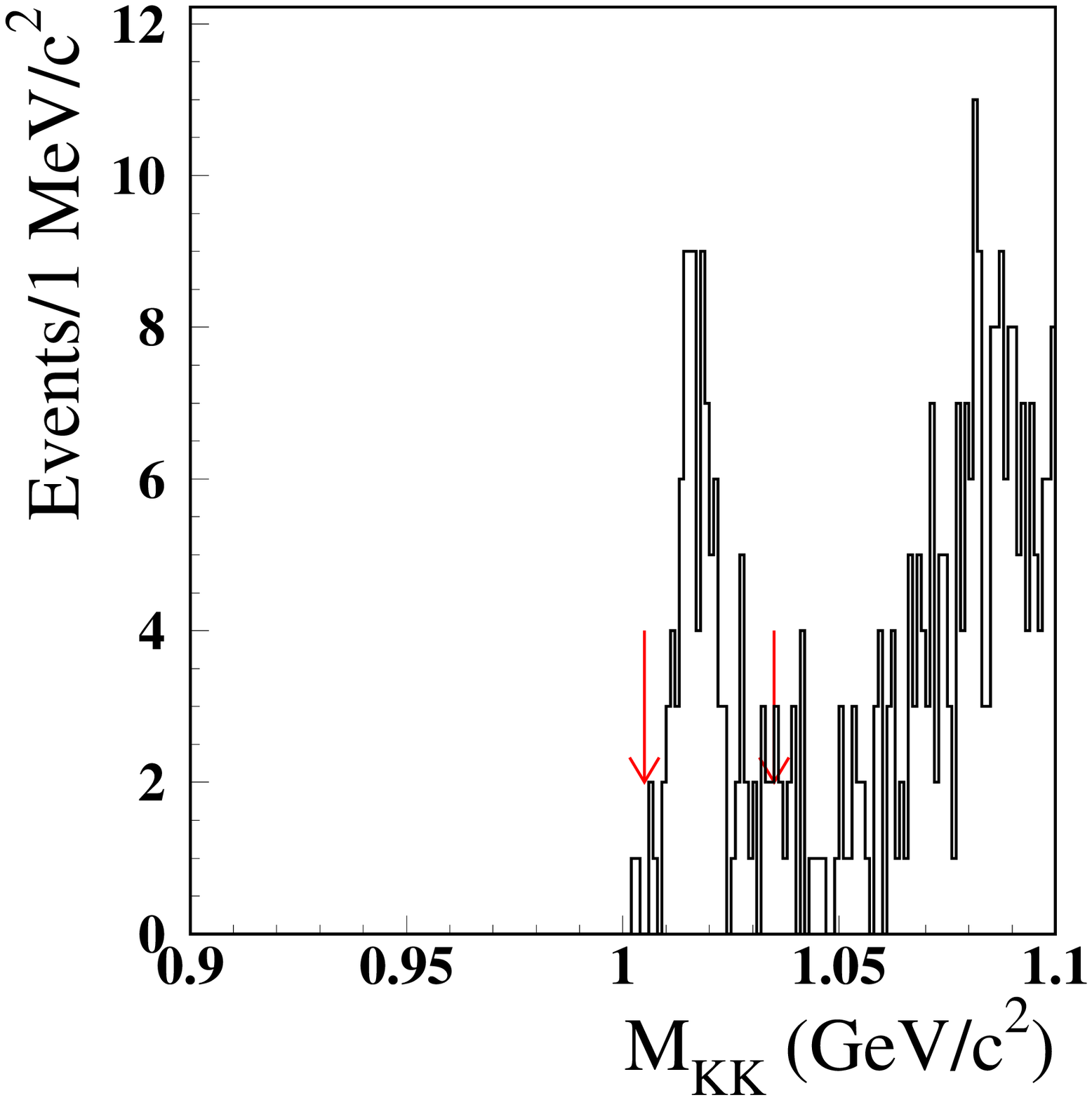}
  \put(-60,100){(a)}
  \end{minipage}
  \begin{minipage}{1.0\linewidth}
  \includegraphics[width=\textwidth,height=0.18\textheight]{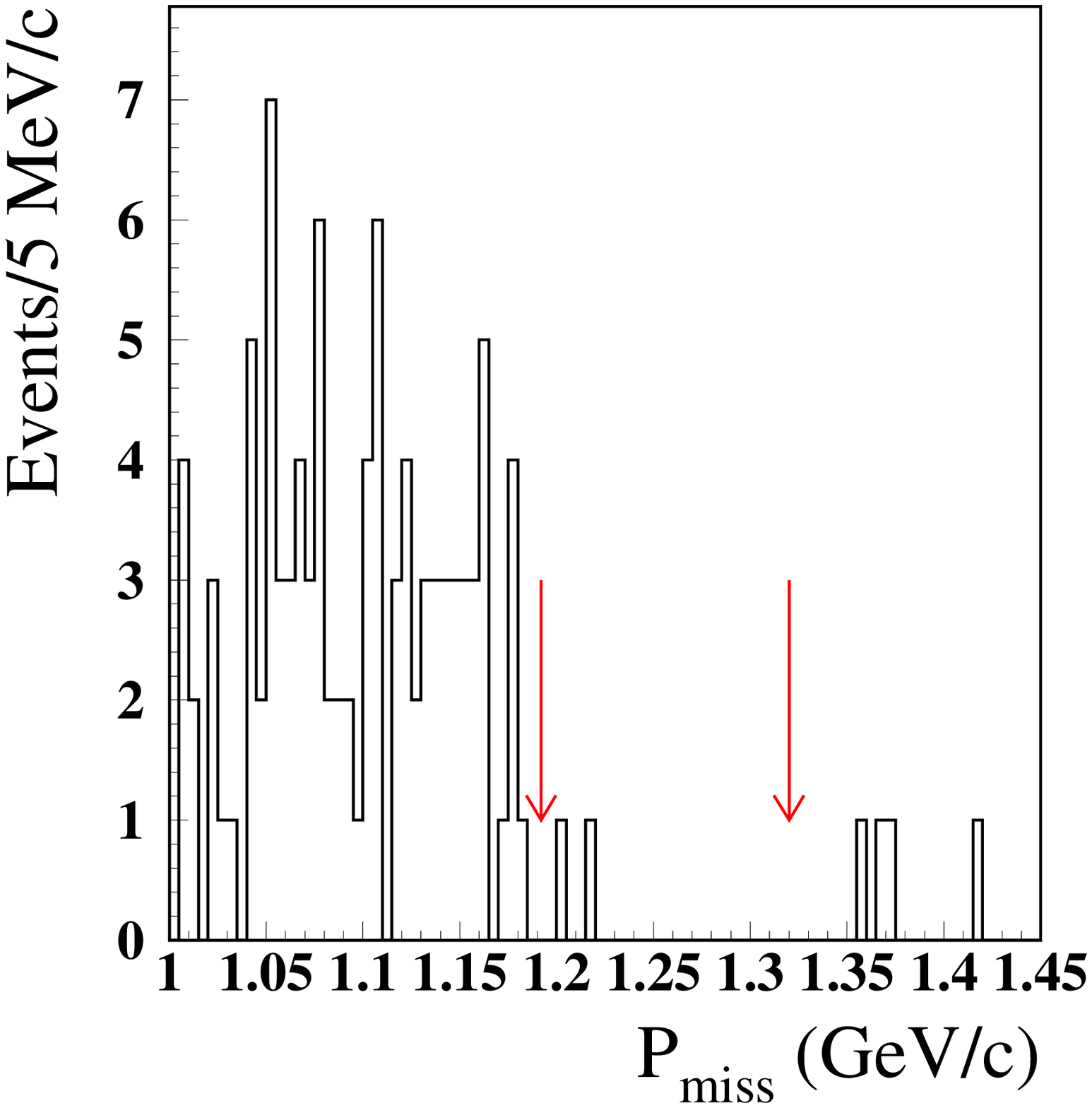}
  \put(-60,100){(b)}
  \end{minipage}
  \caption{(a) The $m_{KK}$ distribution for candidate events.  The
    arrows on the plot indicate the signal region of $\phi$
    candidates.  (b) $P_{miss}$ distribution for the events with
    $1.005<m_{KK}<1.035$ GeV/$c^2$ in (a).  The means of the missing
    momenta for $J/\psi\to\phi\eta$ and $J/\psi\to\phi\eta^\prime$ are
    located around 1.32 and 1.20 GeV/$c$, respectively, as indicated
    by the two arrows.}
  \label{fig:data}
\end{figure}

We require that events have no other charged tracks besides those of
the $\phi \rightarrow K^+K^-$ candidate.  We count the number of BSC
clusters, that could be from a $K^0_L$ or a photon, $N_{BSC}$, and
require that $N_{BSC}$ be zero in the region outside cones of
$30^\circ$ around the charged kaon tracks.  These requirements reject
most $\eta$ and $\eta^{\prime}$ decays into visible final states.
They also eliminate most backgrounds from multi-body decays of $J/\psi
\rightarrow \phi + \text{anything}$.  In order to ensure that $\eta$
and $\eta^{\prime}$ decay particles are inside the fiducial volume of
the detector, the recoil direction against the $\phi$ is required to
be within the region $|\cos\theta_{recoil}|<0.7$, where
$\theta_{recoil}$ is the polar angle of $\vec{P}_{miss}$.
Figure~\ref{fig:data}(a) shows the invariant mass distribution of
$K^+K^-$ candidates, $m_{KK}$, after the above selection. A clear
$\phi$ peak is seen. Figure~\ref{fig:data}(b) shows the $P_{miss}$
distribution for events with $1.005<m_{KK}<1.035$ GeV/$c^2$.

We use Monte Carlo (MC) simulated events to determine selection
efficiencies for the signal channels and study possible backgrounds.
We obtain efficiencies of 23.5\% and 23.2\% for $\eta$ and
$\eta^{\prime}$ invisible decays, respectively.  More than 20
exclusive decay modes are studied with full MC simulations in order to
understand the backgrounds.  The sources of backgrounds are divided
into two classes.  Class I: the background is from J/$\psi \rightarrow
\phi \eta (\eta^{\prime})$, where $\phi \rightarrow K^+K^-$ and $\eta
(\eta^{\prime})$ decays into other modes than the invisible final
states. We find the expected number of background events from this
class is negligible for both $\eta$ and $\eta^{\prime}$. Class II: it
is mainly from $J/\psi$ decays to the final states without $\eta$ or
$\eta^{\prime}$, such as $\phi K_L K_L$, $\phi f_0(980) (f_0(980)
\rightarrow K_L K_L)$, and $K^{\star 0} K_L$($K^{\star 0}\to
K^\pm\pi^\mp$).  For $\eta$ case, the dominated background is from
the decay of $J/\psi \rightarrow K^{\star 0} K_L$($K^{\star 0}\to
K^\pm\pi^\mp$), while for $\eta^\prime$ case, the dominated background is from
the decays of $J/\psi \rightarrow \phi K_L K_L$ and $\phi f_0(980) (f_0(980)
\rightarrow K_L K_L)$.
 The expected number of background events from class
II is $3.0 \pm 0.2$ and $90 \pm 64$ in the signal regions for $\eta$
and $\eta^{\prime}$, respectively.

\begin{figure}[htbp]
  \centering
  \includegraphics[width=0.4\textwidth,height=0.20\textheight]{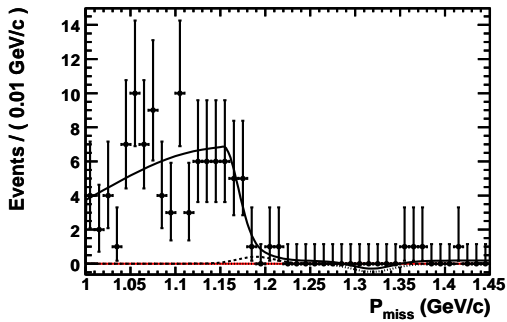}
  \caption{The $P_{miss}$ distribution for candidate events. The data
    (black crosses) are compared to the total fit results. The dotted
    curve is the projection of $\eta$ signal component, and the dashed
    curve is the the projection of $\eta^{\prime}$ signal component,
    and the solid curve is the total likelihood fit result.}
\label{fig:table_eta_etap}
\end{figure}

An unbinned extended maximum likelihood (ML) fit is used to extract
the event yield for J/$\psi \rightarrow \phi\eta (\eta^{\prime})$
[$\phi \rightarrow K^+K^-$ and $\eta(\eta^{\prime}) \rightarrow
\text{invisible}$].  In the ML fit, we require that $1.00
<P_{miss}<1.45$ GeV/$c$, shown in Fig.~\ref{fig:data}(b), where the
background shape is well understood.  We construct probability density
functions (PDFs) for the $P_{miss}$ distributions for
($\mathcal{F}^{\eta}_{sig}$ and $\mathcal{F}^{\eta^{\prime}}_{sig}$)
signals and background ($\mathcal{F}_{bkgd}$) using detailed
simulations of signal and background.  The PDFs for signals are
parameterized by double Gaussian distributions with common means, one
relative fraction and two distinct widths, which are all fixed to the
MC simulation. The PDF for background is a bifurcated Gaussian plus a
first order Polynomial ($P_1$). All parameters related to the
background shape are floated in the fit to data. The PDFs for signals
and background are combined in the likelihood function $\mathcal{L}$,
defined as a function of the free parameters $N^{\eta}_{sig}$,
$N^{\eta^{\prime}}_{sig}$, and $N_{bkgd}$,
\begin{eqnarray}
  \mathcal{L}(N^{\eta}_{sig},N^{\eta^{\prime}}_{sig}, N_{bkgd}) =
  \frac{ e^{-(N^{\eta}_{sig} +
      N^{\eta^{\prime}}_{sig} + N_{bkgd} )} }{N!}  \nonumber\\
  \times \prod^N_{i=1}[N^{\eta}_{sig}\mathcal{F}^{\eta}_{sig}(P^i_{miss})+
  \nonumber \\
   N^{\eta^{\prime}}_{sig}\mathcal{F}^{\eta^{\prime}}_{sig}(P^i_{miss}) +
  N_{bkgd}\mathcal{F}_{bkgd}(P^i_{miss})],
 \label{eq:likelihood}
\end{eqnarray}
where $N^{\eta}_{sig}$ and $N^{\eta^{\prime}}_{sig}$ are the number of
$J/\psi \rightarrow \phi(\rightarrow K^+K^-)\eta(\rightarrow
\text{invisible})$ and $J/\psi \rightarrow \phi(\rightarrow
K^+K^-)\eta^\prime (\rightarrow \text{invisible})$ signal events;
$N_{bkgd}$ is the number of background events.  The fixed parameter
$N$ is the total number of selected events in the fit region, and
$P^i_{miss}$ is the value of $P_{miss}$ for the $i$th event.  The
negative log-likelihood ($- \ln \mathcal{L}$) is then minimized with
respect to $N^{\eta}_{sig}$, $N^{\eta^{\prime}}_{sig}$, and $N_{bkgd}$
in the data sample. A total of 105 events are used in the fit, and the
resulting fitted values of $N^{\eta}_{sig}$,
$N^{\eta^{\prime}}_{sig}$, and $N_{bkgd}$ are $-2.8\pm 1.4$, $2.2 \pm
3.4$, and $106\pm 11$, where the errors are statistical.
Figure~\ref{fig:table_eta_etap} shows the $P_{miss}$ distribution and
fitted result.  No significant signal is observed for the invisible
decay of either $\eta$ or $\eta^\prime$. We obtain upper limits by
integrating the normalized likelihood distribution over the positive
values of the number of signal events. The upper limits at the 90\%
confidence level are 3.56 events for $\eta$ and 5.72 events for
$\eta^\prime$, respectively.

The branching fraction of $\eta (\eta^{\prime})\rightarrow \gamma
\gamma$ is also determined in $J/\psi \rightarrow \phi \eta
(\eta^{\prime})$ decays, in order to obtain the ratio of
$\mathcal{B}(\eta(\eta^\prime) \rightarrow \text{invisible})$ to
$\mathcal{B}(\eta(\eta^\prime) \rightarrow \gamma \gamma)$.  The
advantage of measuring
$\displaystyle\frac{\mathcal{B}(\eta(\eta^\prime) \rightarrow
  \text{invisible})}{\mathcal{B}(\eta(\eta^\prime) \rightarrow \gamma
  \gamma)}$ is that the uncertainties due to the total number of
$J/\psi$ events, tracking efficiency, PID, the number of the charged
tracks, the cut on $m_{KK}$, and residual noise in the BSC cancel.

The selection criteria for the charged tracks are the same as those
for \phett, $\eta(\etap)\to \text{invisible}$ decays.  However, at
least two good photons are required.  A candidate photon must have
hits in the BSC. The number of layers hit must be greater than one,
and the deposited energy in the BSC more than 50 MeV. The angle
between the photon emission direction and the shower development
direction of the neutral track in BSC is required to be less than
$25^\circ$. The opening angles between the candidate photons and the
charged tracks must be greater than $30^\circ$.

\begin{figure}[bp]
  \centering
  \includegraphics[width=0.45\linewidth,height=0.15\textheight]{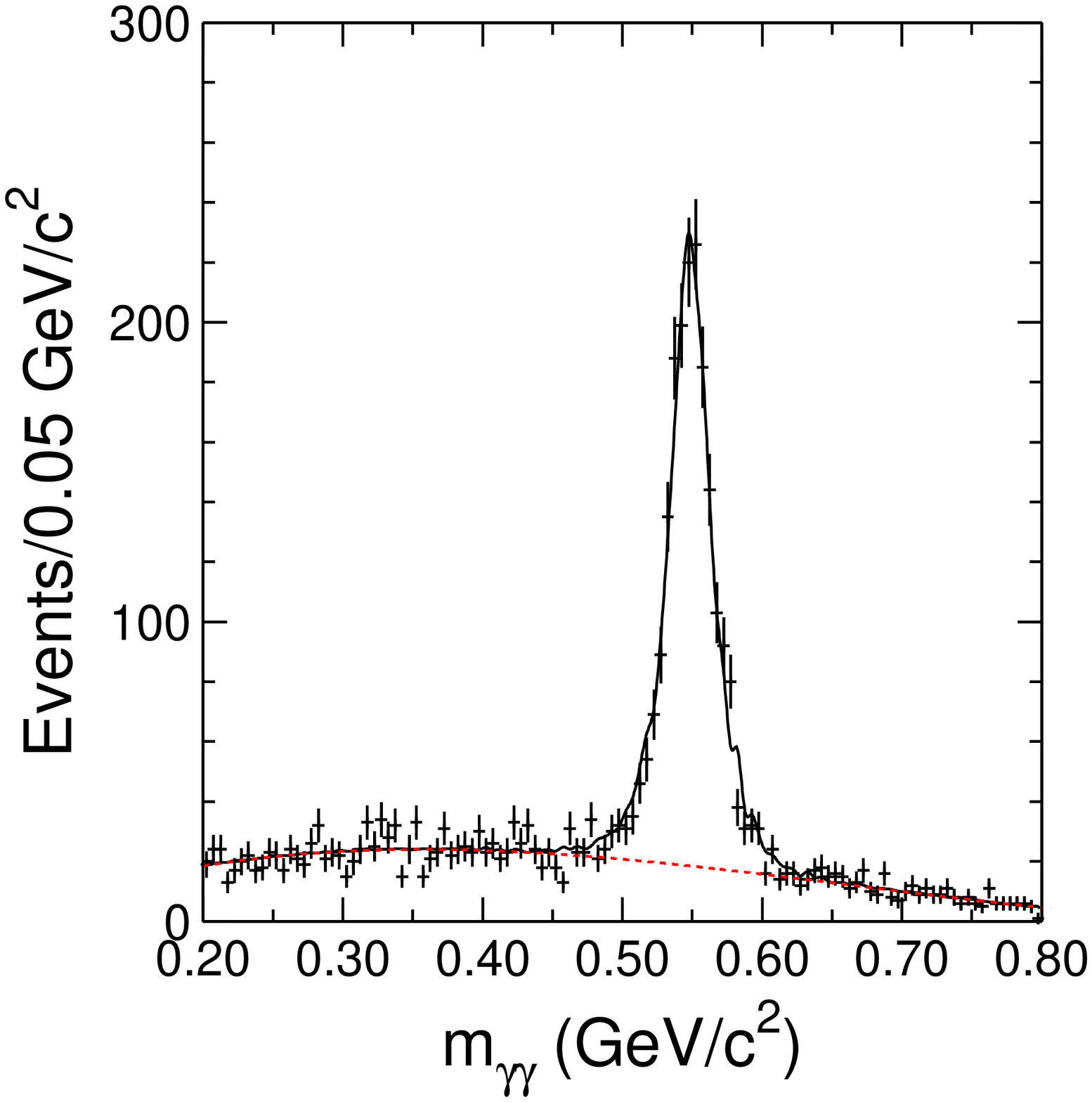}
  \put(-80, 80){(a)}
  \includegraphics[width=0.45\linewidth,height=0.15\textheight]{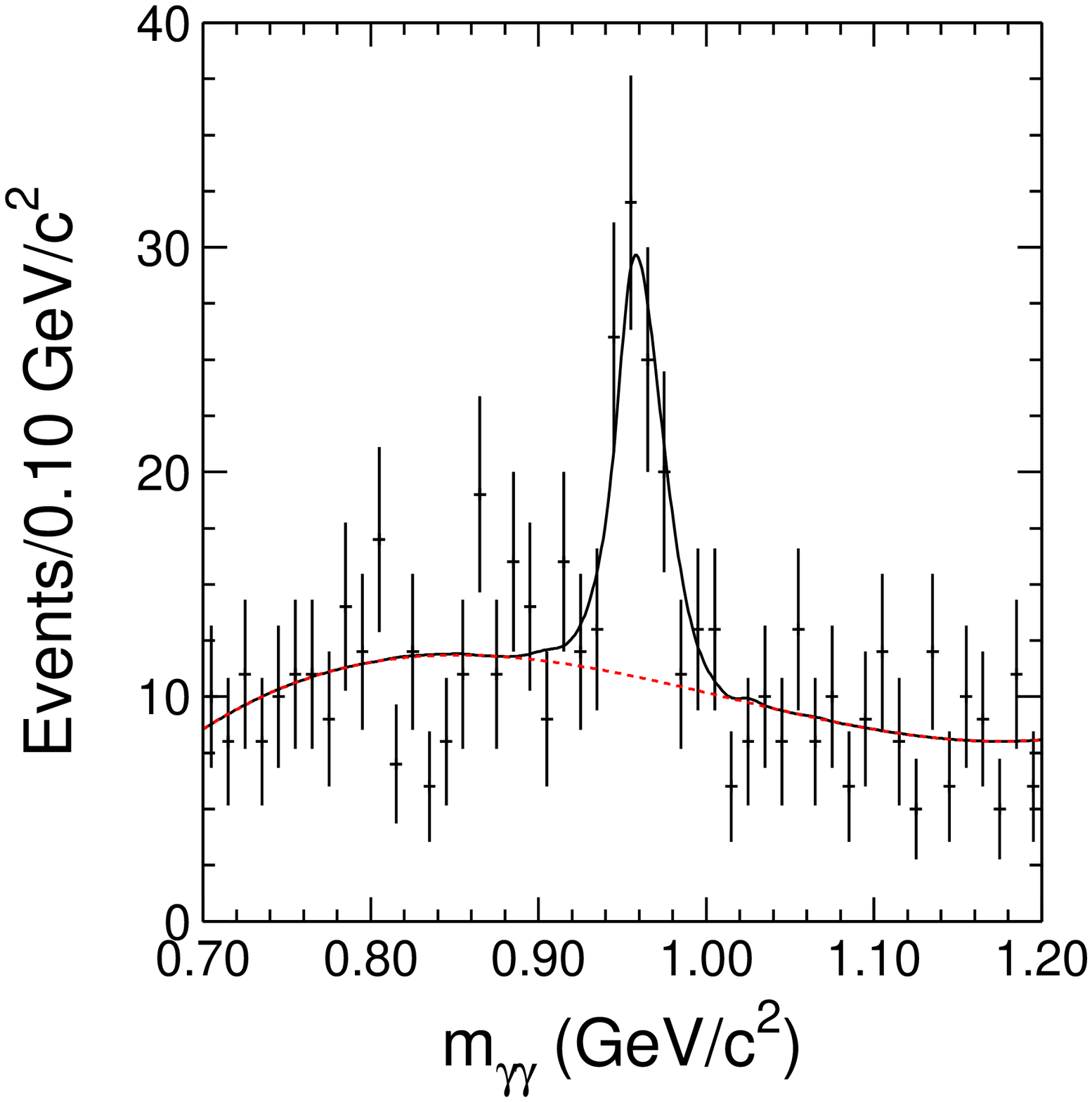}
  \put(-80, 80){(b)}
  \caption{(a) The fit of the $\gamma \gamma$ invariant mass
    distribution of $J/\psi \rightarrow \phi \eta$, $\eta \rightarrow
    \gamma \gamma$.  The dashed line shows the background, and the
    solid line is the total fit result.  (b) The same plot but for
    $J/\psi \rightarrow \phi \eta^\prime$, $\eta^\prime \rightarrow
    \gamma \gamma$. }
  \label{fig:gg}
\end{figure}

The events are kinematically fitted using energy and momentum
conservation constraints (4-C) under the $J/\psi\to KK\gamma\gamma$
hypothesis in order to obtain better mass resolution and suppress
backgrounds further.  We require the kinematic fit
$\chi^2_{K^+K^-\gamma\gamma}$ less that 50 (15) for the
$\eta~(\eta^\prime)$ case. If there are more than two photons, the fit
is repeated using all permutations, and the combination with the best
fit to $K^+K^-\gamma\gamma$ is retained.  The numbers of \phett [$\phi
\rightarrow K^+K^-$ and $\eta(\etap)\to\gamma\gamma$] events are
obtained from fits to the $\gamma\gamma$ invariant mass distributions.
The fitted results for $\eta(\eta^\prime) \rightarrow \gamma \gamma$
are shown in Fig.~\ref{fig:gg}.

Contributions to the systematic error on the ratios are summarized in
Table~\ref{tab:syserr}. Systematic errors in the ML fit originate from
the limited number of events in the data sample and from uncertainties
in the PDF parameterizations.  The uncertainty due to the background
shape has been estimated by varying the PDF shape of the background in
the ML fit.

The uncertainty, due to the requirement of no neutral clusters in the
BSC allowed outside the $30^o$ cones around the charged tracks, is
obtained using the control sample of fully reconstructed $J/\psi
\rightarrow \phi \eta $, $\eta \rightarrow \gamma \gamma$ events.  The
ratios of events with the requirement on the number of extra photons
to events without the requirement are obtained for both data and MC
simulation. The difference, 5\%, is considered as the systematic error
for both the $\eta$ and $\eta^{\prime}$ cases. This study determines
the difference in the noise in the BSC for MC simulation and data.
Compared with $\eta \rightarrow \text{invisible}$ decay, we expect
that more noise is introduced by the photons in $\eta \rightarrow
\gamma \gamma$ decay. So it is a conservative estimation of the
systematic error due to the requirement of no clusters in the BSC for
the invisible decays of $\eta$ and $\eta^\prime$.

The uncertainty in the determination of the number of observed \phett,
$\phi\to K^+ K^-$, $\eta(\etap)\to\gamma\gamma$ events is also
estimated.  Different background shapes are tried in the fit to the
$\gamma\gamma$ invariant mass, and the variation of the fitted yields
is regarded as a systematic error, which is 2.0\% (1.0\%) for the
$\eta$ ($\etap$) case. The relative systematic error caused by the
uncertainty of the photon efficiency is about
4.0\%~\cite{Ablikim:2004hz}. The uncertainty due to the
$\chi^2_{{K^+K^-\gamma\gamma}}$ constraint is estimated to be 1.0\%
(5.2\%)~\cite{Bai:2004jn} for the $\eta$ ($\etap$) case.  The
uncertainty from the trigger efficiency is also considered. The total
systematic error, $\sigma_\eta^{sys}$ ($\sigma_{\eta^\prime}^{sys}$),
on the ratio is 7.7\% (11.1\%) for $\eta$ ($\eta^\prime$), as
summarized in Table~\ref{tab:syserr}.
\begin{table}[htbp]
  \centering
  \caption{Summary of relative systematic errors. The first
    three lines are for
    $J/\psi \rightarrow \phi\eta(\eta^\prime)$,
    $\eta(\eta^\prime)\to \text{invisible}$. The next three
    are for $J/\psi
    \rightarrow \phi \eta(\eta^\prime)$,  $\eta(\eta^\prime)\to\gamma\gamma$.}
  \begin{tabular}{l|cc}
    \hline \hline Source of Uncertainties
    & \multicolumn{2}{|c}{sys. error (\%)} \\
    & $\eta$ & $\etap$ \\ \hline
    PDF shapes in the ML fit  & 3.4 & 7.3 \\
    MC statistics  & 1.0 & 1.0 \\
    Requirement on $N_{BSC}$ & 5.0 & 5.0 \\
    Photon efficiency & 4.0 & 4.0 \\
    4-C fit for $\eta (\eta^\prime) \rightarrow \gamma \gamma$ & 1.0 & 5.2 \\
    Background shape for $\eta (\eta^\prime) \rightarrow \gamma \gamma$ & 2.0 & 1.0  \\ \hline
    Total & 7.7 &11.1 \\ \hline \hline
  \end{tabular}
  \label{tab:syserr}
\end{table}

The upper limit on the ratio of the $\mathcal{B}(\eta\to
\text{invisible})$ to $\mathcal{B}(\eta\to\gamma\gamma)$ is calculated
with
\begin{eqnarray}
  \frac{\mathcal{B}(\eta\to \text{invisible})}{\mathcal{B}(\eta\to\gamma\gamma)} <
  \frac{n^{\eta}_{UL}/\epsilon_\eta}{n^{\eta}_{\gamma\gamma}/\epsilon^\eta_{\gamma\gamma}}\cdot
  \frac{1}{(1-\sigma_\eta)}
  \label{eq:upper}
\end{eqnarray}
where $n^{\eta}_{UL}$ is the 90\% upper limit of the number of
observed events for \phet, $\phi\to K^+K^-$, $\eta\to
\text{invisible}$ decay, $\epsilon_{\eta}$ is the MC determined
efficiency for the signal channel, $n^\eta_{\gamma\gamma}$ is the
number of events for the \phet, $\phi\to K^+K^-$,
$\eta\to\gamma\gamma$ decay, $\epsilon^\eta_{\gamma\gamma}$ is the MC
determined efficiency for the decay mode, and $\sigma_{\eta}$ is
$\sqrt{(\sigma^{sys}_{\eta})^2+(\sigma^{stat}_{\eta})^2} = 8.1\%$,
where $\sigma^{sys}_{\eta}$ and $\sigma^{stat}_{\eta}$ are the total
relative systematical error for the $\eta$ case from
Table~\ref{tab:syserr} and the relative statistical error of
$n^\eta_{\gamma\gamma}$, respectively. For $\eta^\prime$,
$\sigma_{\eta^\prime}$ is
$\sqrt{(\sigma^{sys}_{\eta^\prime})^2+(\sigma^{stat}_{\eta^\prime})^2}=21.6\%$.
The relative statistical error of the fitted yield for $J/\psi
\rightarrow \phi \eta(\eta^\prime)$,
$\eta(\eta^\prime)\to\gamma\gamma$, is 2.8\% (18.5\%) according to the
results from the fit to the invariant mass of $\gamma \gamma$ in
Fig.~\ref{fig:gg}.  We also obtain the upper limit on the ratio of the
$\mathcal{B}(\eta^\prime \to \text{invisible})$ to
$\mathcal{B}(\eta^\prime\to\gamma\gamma)$ by replacing $\eta$ with
$\eta^\prime$ in Eq.~(\ref{eq:upper}).  Since only the statistical
error is considered when we obtain the 90\% upper limit of the
number of events, to be conservative, $n^{\eta}_{UL}$ and
$n^{\eta^\prime}_{UL}$ are shifted up by one sigma of the
additional uncertainties ($\sigma_{\eta}$ or
$\sigma_{\eta^\prime}$).
\begin{table}[htbp]
  \centering
  \caption{The numbers used in the calculations of the ratios in
    Eq.~(\ref{eq:upper}), where $n^\eta_{UL}~(n^{\eta^\prime}_{UL})$ is
    the upper limit of the signal events at the 90\% confidence level,
    $\epsilon_\eta~(\epsilon_{\eta^\prime})$ is the selection
    efficiency,
    $n^\eta_{\gamma\gamma}~(n^{\eta^\prime}_{\gamma\gamma})$ is the
    number of the events of $J/\psi\to\phi\eta(\eta^\prime)$, $\phi\to
    K^+ K^-$, $\eta(\eta^\prime)\to\gamma\gamma$,
    $\epsilon^\eta_{\gamma\gamma}~(\epsilon^{\etap}_{\gamma\gamma})$ is
    its selection efficiency, $\sigma_\eta^{stat}$
    ($\sigma_{\eta^\prime}^{stat}$) is the relative statistical error
    of $n^\eta_{\gamma\gamma}~(n^{\eta^\prime}_{\gamma\gamma})$ and
    $\sigma_\eta~(\sigma_{\eta^\prime})$ is the total relative error.}
  \begin{tabular}{l|cc}
    \hline \hline
    quantity & \multicolumn{2}{|c}{value} \\
    & $\eta$ & $\etap$ \\\hline
    $n^{\eta}_{UL}$ ($n^{\eta^\prime}_{UL}$) & 3.56 & 5.72 \\
    $\epsilon_\eta$ ($\epsilon_{\eta^\prime}$) & $23.5$\% & $23.2$\% \\
    $n^{\eta}_{\gamma\gamma}$ ($n^{\eta^\prime}_{\gamma\gamma}$) & $1760.2 \pm 49.3$ & $71.6 \pm 13.2$ \\
    $\epsilon^\eta_{\gamma\gamma}$ ($\epsilon^{\etap}_{\gamma\gamma}$) & $17.6$\% & $15.2$\% \\
    $\sigma_\eta^{stat}$ ($\sigma_{\eta^\prime}^{stat}$) & 2.8\% & 18.5\% \\
    $\sigma_\eta$ ($\sigma_{\eta^\prime}$) & 8.1\% & 21.6\% \\
    \hline\hline
  \end{tabular}
  \label{tab:br}
\end{table}

Using the numbers in Table~\ref{tab:br}, the upper limit on the ratio
of ${\mathcal B} (\eta(\etap) \rightarrow \text{invisible})$ and
${\mathcal B}(\eta(\etap) \rightarrow \gamma \gamma)$ is obtained at
the 90\% confidence level of $1.65\times 10^{-3}$ ($6.69\times
10^{-2}$).

In summary, we search for the invisible decay modes of $\eta$ and
$\eta^\prime$ for the first time in \phett \mbox{} using the 58
million $J/\psi$ events at BES II. We find no signal yields for the
invisible decays of $\eta$ and $\eta^\prime$, and obtain limits on the
ratio, $\displaystyle\frac{\mathcal{B}(\eta(\eta^\prime) \to
  \text{invisible})}{\mathcal{B}(\eta(\eta^\prime) \to\gamma\gamma)}$.
The upper limits at the 90\% confidence level are $1.65\times 10^{-3}$
and $6.69\times 10^{-2}$ for $\displaystyle\frac{\mathcal{B}(\eta \to
  \text{invisible})}{\mathcal{B}(\eta \to\gamma\gamma)}$ and
$\displaystyle\frac{\mathcal{B}(\eta^\prime \to
  \text{invisible})}{\mathcal{B}(\eta^\prime \to\gamma\gamma)}$,
respectively.  The advantage of measuring the ratios instead of the
branching fractions of the invisible decays is that many uncertainties
cancel.

\begin{acknowledgements}
  The BES collaboration thanks the staff of BEPC and computing center
  for their hard efforts. We also thank Pierre Fayet for illuminating
  suggestions. This work is supported in part by the National Natural
  Science Foundation of China under contracts Nos.  10491300,
  10225524, 10225525, 10425523, the Chinese Academy of Sciences under
  contract No. KJ 95T-03, the 100 Talents Program of CAS under
  Contract Nos. U-11, U-24, U-25, and the Knowledge Innovation Project
  of CAS under Contract Nos. U-602, U-612, U-34 (IHEP), the National
  Natural Science Foundation of China under Contract No.  10225522
  (Tsinghua University), and the Department of Energy under Contract
  No.DE-FG02-04ER41291 (U Hawaii).
\end{acknowledgements}

\end{document}